\documentstyle[amssymb,12pt,psfig,epsfig]{article}

\def\be{\begin{equation}}
\def\ee{\end{equation}}
\def\ba{\begin{eqnarray}}
\def\ea{\end{eqnarray}}
\def\ga{\mathrel{\raise.3ex\hbox{$>$\kern-.75em\lower1ex\hbox{$\sim$}}}}
\def\la{\mathrel{\raise.3ex\hbox{$<$\kern-.75em\lower1ex\hbox{$\sim$}}}}

\def\simgt{\mathrel{\raise.3ex\hbox{$>$\kern-.75em\lower1ex\hbox{$\sim$}}}}
\def\simlt{\mathrel{\raise.3ex\hbox{$<$\kern-.75em\lower1ex\hbox{$\sim$}}}}

\newcommand{\fr}[2]{\frac{#1}{#2}}

\newcommand{\nc}{\newcommand}

\nc{\gone}{\bar g_{\pi NN}^{(1)}}
\nc{\gzero}{\bar g_{\pi NN}^{(0)}}
\nc{\al}{\alpha}
\nc{\de}{\delta}
\nc{\ep}{\epsilon}
\nc{\ze}{\zeta}
\nc{\et}{\eta}
\renewcommand{\th}{\theta}
\nc{\Th}{\Theta}
\nc{\ka}{\kappa}
\nc{\rh}{\rho}
\nc{\si}{\sigma}
\nc{\ta}{\tau}
\nc{\up}{\upsilon}
\nc{\ph}{\phi}
\nc{\ch}{\chi}
\nc{\ps}{\psi}
\nc{\om}{\omega}
\nc{\Ga}{\Gamma}
\nc{\De}{\Delta}
\nc{\La}{\Lambda}
\nc{\Si}{\Sigma}
\nc{\Up}{\Upsilon}
\nc{\Ph}{\Phi}
\nc{\Ps}{\Psi}
\nc{\Om}{\Omega}
\nc{\ptl}{\partial}
\nc{\del}{\nabla}
\nc{\ov}{\overline}
\input{tcilatex}
\begin{document}

\begin{titlepage}
\rightline{UVIC--TH--04/01}
\rightline{hep-ph/yymmnn}
\rightline{February 2004}
\begin{center}

\large {\bf Unstable Relics as a Source of Galactic Positrons}\\[3mm]

\normalsize

{\bf Charles Picciotto$^{\,(a)}$} and  { {\bf Maxim Pospelov$^{\,(a,b)}$} }

\smallskip
\medskip

$^{\,(a)}${\it Department of Physics and Astronomy, University of Victoria, \\
     Victoria, BC, V8P 1A1 Canada}

$^{\,(b)}${\it Centre for Theoretical Physics,
                                 CPES,
                                 University of Sussex,
                                 Brighton BN1~9QJ,\\
                                 United Kingdom }

\smallskip
\end{center}
\vskip0.6in

\centerline{\large\bf Abstract}

We calculate the fluxes of 511 KeV photons from the Galactic 
bulge caused by positrons 
produced in the decays of relic particles with masses less than 
100 MeV. In particular, 
we tighten the constraints on sterile neutrinos over a large domain of the 
mass--mixing angle parameter space, where the resulting photon flux would significantly 
exceed the experimental data. At the same time, the observed 
photon fluxes can be easily caused 
by decaying sterile neutrinos in the mass range $1$ MeV$<m_{\rm sterile} \la 
50 $ MeV with the cosmological abundance typically within 
$10^{-9}\la \Omega_{\rm sterile} \la 10^{-5}$, assuming 
that $\Omega_{\rm sterile} $ comes entirely from the conversion of 
active neutrinos in the early Universe. 
Other candidates for decaying relics such as neutral (pseudo)scalar particles
coupled to leptons with the gravitational strength can be compatible with the 
photon flux, and can constitute the main component of cold dark matter. 

\vspace*{2mm}

\end{titlepage}

\section{Introduction}

Recent observations of 511 KeV photons from the galactic center produced by
positron annihilation performed by the SPI spectrometer on the space
observatory INTEGRAL \cite{SPI} confirm previously reported photon fluxes 
\cite{positrons} and improve the angular resolution. New data are
inconsistent with a single point source, although more data with better
angular resolution are needed to exclude or resolve multiple point sources.
An apparent de-localization of the source around the galactic center
combined with the absence of well-recognized astrophysical sources of
positrons that can fill the galactic bulge have prompted speculations about
a non-Standard Model  origin of galactic positrons.

An interesting explanation was suggested recently in Ref. \cite{Silk}, where
galactic positrons were attributed to annihilating dark matter. In order to
have photon fluxes at the observable level, one needs rather large number
densities of dark matter particles near the center that requires smaller
masses than the 1GeV - 1 TeV range preferred by most models of weakly
interacting massive particles. Combined with the additional requirement that
the produced positrons slow down to non-relativistic velocities before
annihilation, this puts dark matter particles in a mass range of 1 MeV to
100 MeV \cite{Silk}. According to the scenario of Ref. \cite{Silk}, the 
galactic positrons can be effectly trapped by the magnetic field and 
dissipate their energy by multiple collisions with the gas. The stopping distance 
is estimated to be comparable or shorter than the "free path" before the annihilation
for positrons with energies less than about 100 MeV.   
An interesting spin-off of this scenario is the
possibility to probe the steepness of the dark matter halo profile near the
center of the galaxy. Using the power-like parametrization (see e.g. \cite
{NFW}) of dark matter energy density, $\rho _{DM}\sim r^{-\gamma }$, near
the center, the Oxford-Paris group \cite{Silk} concluded that $0.4%
\mathrel{\raise.3ex\hbox{$<$\kern-.75em\lower1ex\hbox{$\sim$}}}\gamma %
\mathrel{\raise.3ex\hbox{$<$\kern-.75em\lower1ex\hbox{$\sim$}}}0.8$ is
preferred. With more data coming in the future with better angular
resolution, and with the better understanding of conventional (dark-matter
unrelated) sources of non-relativistic positrons, the flux of 511 KeV
photons can provide more detailed information on the distribution of dark
matter near the center of our galaxy. Even though the explanation of the 511
KeV line from the bulge by annihilating dark matter certainly deserves
further considerations, one can easily see that this requires very unusual
properties of dark matter. Smaller mass and relatively large annihilation
cross sections to electron-positron pairs would require either extreme fine
tuning in the models of dark matter (see e.g. \cite{BPtV}) or introduction
of new mediating forces that would enable more efficient annihilation \cite
{Fayet}.

In this paper, we argue that the decay of quasi-stable relic particles can
provide sufficient amount of positrons to explain the SPI/INTEGRAL data.
These relics may be the main component of the dark matter, or a sub-dominant
fraction of it. From the point of view of properties of dark matter, this
would not be as restrictive as the scenario with annihilating dark matter.
At the same time, it is clear that if the decaying dark matter can provide
sufficient numbers of positrons, the resulting photon angular distribution would
imply {\em steeper} profiles of dark matter near the center than in the
scenario with annihilating dark matter.

To substantiate our claims we consider two types of relics that can easily
provide a sufficient amount of positrons. Sterile neutrinos in the mass range
of few KeV - few MeV have been extensively studied in a variety of
cosmological and astrophysical settings. Their cosmological abundance is
thoroughly investigated by now as a function of their mass $m_{s}$, mixing
angle $\th $ with active neutrinos \cite{DW}, as well as their lepton asymmetry and 
flavor \cite{DH,AFP}. We calculate the decay width of sterile
neutrinos and their branching to positrons which is always larger than the
branching ratio to gamma quanta if the decay into $e^{+}e^{-}$ pair is
kinematically allowed, $m_{s}>$1 MeV. We demonstrate that in the large part
of the $\th -m_{s}$ parameter space the resulting flux of photons is {\em %
too large} to be compatible with observations, and therefore is excluded. 
Thus, the observed fluxes of 511 KeV photons \cite{positrons}
firmly rule out sterile neutrino dark matter with $m_{s}>$1 MeV.
On the fringe of this parameter space lies an allowed area compatible with the
SPI/INTEGRAL observations. This area is large enough to span a wide range of
the parameter space, compatible with sterile neutrino abundances $10^{-9} %
\mathrel{\raise.3ex\hbox{$<$\kern-.75em\lower1ex\hbox{$\sim$}}}\Omega _{{\rm %
sterile}}\mathrel{\raise.3ex\hbox{$<$\kern-.75em\lower1ex\hbox{$\sim$}}}
10^{-5}$.  We also show that the constraints on the parameter space of the model 
from the positron annihilation in the galactic center are complimentary to those 
coming from the excessive cosmological background of gamma quanta 
produced in the decay of sterile neutrinos. 

Other models of unstable relics can be more flexible, and accommodate dark
matter and observable fluxes of photons at the same time. Here we consider
models with scalar and/or pseudoscalar particles with MeV range masses and 
$O(M_{{\rm Pl}}^{-1})$ couplings to matter. Such scalar particles can be
motivated by models with extra space-like dimensions, where radions may have
similar properties. We argue the electron-positron channel can be a dominant
decay mode for these scalars and that they can be the main component of 
cold dark matter.

\section{Sterile neutrinos as a source of positrons}

Sterile neutrinos are the simplest and the most credible extension of the
Standard Model. Right-handed neutrinos with a relatively heavy Majorana mass 
$m_{R}$ are often used to explain the apparent smallness of the mass range for
the left-handed neutrinos hinted by various neutrino experiments. A priori,
the right-handed mass $m_{R}$ can be in a rather large mass range, from the
GUT scale all the way down to the KeV scale.

In this paper, we consider a model of a sterile neutrino that is
characterized by two main parameters, the mass $m_{s}$ and the mixing angle $%
\theta $ to the left-handed neutrino: 
\begin{equation}
N=\cos \theta N_{R}+\sin \theta \nu _{L};\;\;\nu =\cos \theta \nu _{L}-\sin
\theta N_{R},  \label{decomp}
\end{equation}
where $N_{R}$ refers to a pure SM singlet. We do not confine our discussion
to the see-saw model and allow $\theta $ and $m_{s}$ to vary independently,
without imposing a $m_{\nu }\sim \theta ^{2}m_{s}$ relation. Such models
were studied extensively in the mass range of few KeV - few MeV 
\cite{DW,DHRS,DH,AFP,Irina}, where sterile neutrinos could constitute entire cold dark matter or at
least an appreciable portion of it. Note that the left-handed neutrino in (%
\ref{decomp}) need not be a mass eigenstate, and therefore the sterile
neutrino $N$ can have interaction with all three charged leptons. Thus, the
decay of $N$ to positrons will have an additional parameter $V$, $|V|<1$.
With this parameter, the interaction of sterile neutrinos with a light
active neutrino and electrons takes the following form: 
\begin{equation}
{\cal L}_{{\rm int}}=\frac{\sin 2\th }{2}\left[ \frac{eZ_{\mu }}{2\sin \th
_{W}\cos \th _{W}}\bar{N}\gamma _{\mu }\nu +\frac{g_{w}VW_{\mu }^{+}}{\sqrt{2%
}}\bar{N}\gamma _{\mu }e+\frac{g_{w}V^{*}W_{\mu }^{-}}{\sqrt{2}}\bar{e}%
\gamma _{\mu }N\right] +{\rm higgs~terms}.  \label{wc}
\end{equation}
Note that the existence of interactions with Higgs boson(s) may be important
for the cosmological production of $N$ only if $m_{s}\mathrel{\raise.3ex%
\hbox{$>$\kern-.75em\lower1ex\hbox{$\sim$}}}$1 GeV, which is outside of the
range of 1 MeV - 100 MeV that we are interested in here. Thus, we neglect
Higgs interactions in any further analysis.

In the mass range of 1 MeV - 100 MeV, the sterile neutrino $N$ can decay
into three light neutrinos, light neutrino and a $e^{+}e^{-}$ pair, and a
neutrino and $\gamma $. The decay widths depend on the nature of $N$, which
can be both Dirac or Majorana. For a Majorana particle, the decay width is
twice the decay rate for a Dirac particle. This is simply because two
decay channels, $N\to \nu \nu \bar{\nu}$ and $N\to \nu \bar{\nu}\bar{\nu}$,
are allowed for a Majorana neutrino. For the Dirac sterile neutrino $N$ the
decay width into three light active neutrinos (all three flavors) is given
by 
\begin{equation}
\Gamma _{N\to \nu \nu \bar{\nu}}=\Gamma _{0}\frac{\sin ^{2}2\theta }{4}%
\left( \frac{m_{s}}{m_{\mu }}\right) ^{5},  \label{3nu}
\end{equation}
where $\Gamma _{0}$ is the muon decay width, $\Gamma _{0}=(192\pi
^{3})^{-1}G_{F}^{2}m_{\mu }^{5}$, introduced for convenience.

The decay of $N$ into one light neutrino and the electron-positron pair may
occur due to charged and neutral currents. To simplify our formulae, we
assume to a few per cent accuracy $1-4\sin ^{2}\th _{W}\simeq 0,$ or $\sin
^{2}\th _{W}\simeq 0.25$. In this approximation, and in the limit of $%
m_{e}/m_{s}\ll 1$, the decay width to $\nu e^{+}e^{-}$ summed over all
neutrino flavors can be expressed as 
\begin{equation}
\Gamma _{N\to \nu e\bar{e}}=\Gamma _{0}\frac{\sin ^{2}2\theta }{4}\left( 
\frac{m_{s}}{m_{\mu }}\right) ^{5}\left[ \frac{|V|^{2}}{2}+\frac{1}{8}%
\right] .  \label{e+e-}
\end{equation}
More accurate formulae that take into account the phase space suppression
for $m_{s}\sim 2m_{e}$ and more accurate values of $\sin ^{2}\th _{W}$ can
be easily derived but not needed for the present discussion. The complete 
result can be found in Ref. \cite{BPS}. 
In Eq. (\ref{e+e-}) we take into account $Z$-exchange and its interference with 
$W$-exchange for an arbitrary $V$. The existence of both charged 
and neutral currents is important, because
even for $V=0$, $\Gamma _{N\to \nu e\bar{e}}$ is different from zero, which
ensures a non-zero positron width even for ''muon'' or ''tau'' sterile
neutrinos.

The decay width into $\nu \gamma $ \cite{BPS,WP} contributes less than 1\% to
the total decay width, 
\begin{equation}
\Gamma _{N\to \nu \gamma }=\Gamma _{0}\frac{\sin ^{2}2\theta }{4}\left( 
\frac{m_{s}}{m_{\mu }}\right) ^{5}\frac{27\alpha }{8\pi },
\end{equation}
so that the total decay width is given by the sum of $\nu e^{+}e^{-}$ and $%
\nu \nu \bar{\nu}$ channels, $\Gamma _{N}\simeq \Gamma _{N\to \nu \nu \bar{%
\nu}}+\Gamma _{N\to \nu e\bar{e}}$. The branching ratio to $e^{+}e^{-}$ has
a very simple form, and depends only on the size of the mixing matrix
element $V$: 
\begin{equation}
{\rm Br}_{e^{+}e^{-}}=\frac{\Gamma _{N\to \nu e\bar{e}}}{\Gamma _{N}}=\frac{%
1+4|V|^{2}}{9+4|V|^{2}}.
\end{equation}
When $|V|^{2}=1$ the branching ratio to $e^{+}e^{-}$ reaches the maximum
value of 5/13, while $V=0$ minimizes it, ${\rm Br}_{e^{+}e^{-}}^{{\rm min}%
}=1/9$. Thus, the intrinsic model dependence of ${\rm Br}_{e^{+}e^{-}}$ is a
factor of 3.5.

Having determined the total decay width of $N$, we are ready to approximate
the cosmological abundance of sterile neutrinos by combining the exponential
decay factor with the power-like pre-exponent reflecting the production of
sterile neutrinos in the early Universe. Assuming a lepton asymmetry less
than $10^{-4}$, and using the results of Refs. \cite{DW,AFP}, we arrive at 
\begin{equation}
\Omega _{s}h^{2}=\Omega_{0s}\{-\Gamma _{N}\tau \}
\sim 0.3\left( \frac{\sin ^{2}2\theta }{10^{-14}}\right)
\left( \frac{m_{s}}{10~{\rm MeV}}\right) ^{2}\exp \{-\Gamma _{N}\tau \},
\label{omega}
\end{equation}
where $\tau $ is the age of the Universe. In this formula, $\Omega_{0s}$ is a 
fictitious energy density in neglection of neutrino decay, 
introduced here for future convenience. It coincides with an actual energy density as long 
as $\Gamma _{N}\tau$ is small.  Note that the pre-exponent in Eq. (%
\ref{omega}) is very approximate. In practice, there will be additional
dependence on the lepton asymmetry of the Universe and on elements of the $V$
matrix \cite{AFP}. There is an additional limitation of (\ref{omega}),
namely it works as long as $m_{s}$ is smaller than $T\sim 130$ MeV, that is
the temperature in the early Universe where most of the sterile neutrino
production takes place. 

In the mass range between 1 MeV and 100 MeV sterile neutrinos behave as cold
dark matter. Neutrinos $N$ will be distributed inside a galaxy with the very
same profile as the rest of the cold dark matter. Thus, the number density
of sterile neutrinos as the function of coordinates can be determined as 
\begin{equation}
n_s(r)= \frac{\Omega_s}{\Omega_{DM}} \frac{\rho_{DM}(r)}{m_s},
\label{numden}
\end{equation}
where $\Omega_{DM} = 0.23$ is the "concordance" value for the cold dark matter
energy density, and $\rho_{DM}(r)$ is the cold dark matter energy density in
our galaxy.

Using the above equations, we are ready to estimate the emissivity of
photons that originate from the galactic center. Assuming that all positrons
quickly slow down in matter and annihilate at rest, we estimate the number
of emitted 511 KeV photons per volume per time, 
\begin{equation}
N_{511\gamma }(r)=2n_{s}(r)\Gamma _{N}{\rm Br}_{e^{+}e^{-}},  \label{ngamma}
\end{equation}
from which we can estimate the flux that reaches the Solar system,
\begin{equation}
\Phi _{511\gamma }(\theta )=\frac{1}{4\pi }\int_{{\rm los}}N_{511\gamma
}(r(s))ds,
\end{equation}
where $\theta \,$is the angle from the galactic center and the integration
is along the line of sight. The total flux is obtained by integrating over
the acceptance of the spectrometer. Combining the above equations, one
arrives at the following expression: 
\begin{eqnarray}
\Phi &\simeq &8.7\times 10^{-3}\left( \frac{\sin ^{2}2\theta }{10^{-15}}%
\right) ^{2}\left( \frac{m_{s}}{{\rm MeV}}\right) ^{6}(4|V|^{2}+1)  \nonumber
\\
&&\times \exp \left\{ -5\times 10^{-4}\left( \frac{\sin ^{2}2\theta }{%
10^{-15}}\right) \left( \frac{m_{s}}{{\rm MeV}}\right) ^{5}\left(
4|V|^{2}+9\right) \right\}  \label{flux} \\
&&\times \frac{1}{2\pi }\int d\Omega \int_{{\rm {los }}}\left( \frac{\rho
_{DM}}{{\rm {MeV/cm}^{3}}}\right) d\left( \frac{s}{{\rm {kpc}}}\right)
\,\,\,\,\,{\rm {cm}^{-2}{s}^{-1}.}  \nonumber
\end{eqnarray}

\begin{figure}[tbp]
\hspace*{-2cm} \centerline{\psfig{file=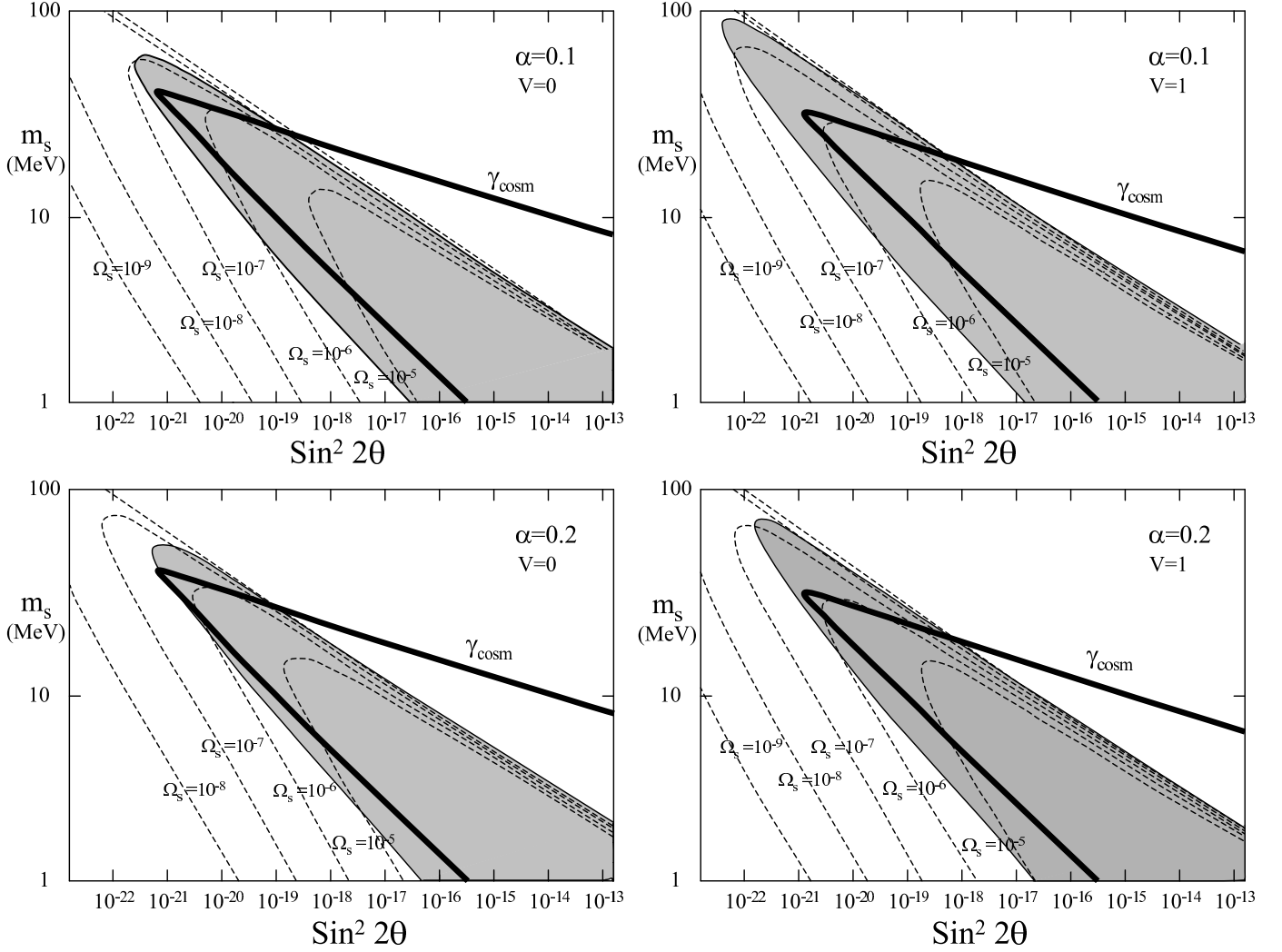,width=12cm,angle=-90}}
\par
\vspace{2.5cm}
\caption{Mixing angle -- sterile neutrino mass parameter space with the
contours of sterile neutrino abundance for different halo profiles, $\alpha
= 0.1$ and 0.2, and for different mixing angle $|V| = 0$ and 1. The dark shaded
region covers the combination of parameters that creates {\em too large} a
flux of 511 KeV photons. The boundary of the shaded region gives the same
flux of photons, as observed by INTEGRAL/SPI. The area  bounded by 
a thick colid curve is excluded by the diffuse cosmological background of gamma photons.}
\end{figure}

Integrals involving the dark matter density are commonly performed with a
NFW density function \cite{NFW} which can be simplified to $\rho
(r)\varpropto 1/r^{\gamma }$ for the inner region of the galaxy in which we
are interested. We have done our calculations for different values of $%
\gamma $. However, a better functional form for the NFW density has been
developed recently \cite{Navarro} which avoids the singularity and thus does
a more careful job of dealing with the radial dependence near the center.
This density function is given by: 
\begin{equation}
\rho _{DM}(r)=\rho _{o}\exp \left\{ -\frac{2}{\alpha }\left[ \left( \frac{r}{%
r_{o}}\right) ^{\alpha }-1\right] \right\} ,  \label{newprof}
\end{equation}
where $r_{o}=20h^{-1}$kpc. To fix an overall normalization $\rho _{o}$ we
take $\rho =0.3$ GeV/cm$^{3}$ at $r=8.5$ kpc, the most commonly used value
for the dark matter energy density near the Solar system. The values of $%
\alpha $ which make this expression consistent with previous density
distributions for larger values of $r$ range between $\alpha =\,0.1$ and $%
\alpha =\,0.2$.

The spectrum of the 511 KeV line emission from the galactic center region
was obtained by the spectrometer SPI on the INTEGRAL gamma-ray observatory.
The data suggest an azimuthally symmetric galactic bulge component with FWHM
of $\sim 9^{{\rm {o }}}$ with a $2\sigma $ uncertainty range covering 6$^{%
{\rm {o }}}$-18$^{{\rm {o}}}$. The flux of 511 KeV photons was measured as $%
\Phi_{{\rm exp}} = 9.9_{-2.1}^{+4.7}\times 10^{-4}$ ph cm$^{-2}$ s$^{-1}$.
We have obtained angular distributions and total fluxes expected from the
decay of sterile neutrinos using the density function (\ref{newprof}) in the
expression for the flux (\ref{flux}) and averaging over the 2$^{{\rm {o}}}$
angular resolution of the spectrometer. By comparing our calculated flux to
the experimental value we find the effective constraints on $\sin
^{2}2\theta $ and $m_{s}$. Fig. 1 shows the results of our calculations for $%
\alpha =0.1$ and $0.2$, for the extreme values of $|V|=0,1$, with the shaded
area providing too large a flux compared to the experimental value.
Therefore, we can exclude a significant portion of the parameter space in
the mass interval $1$ MeV - 100 MeV {\em regardless} what the source of the
galactic positrons really is, conventional astrophysics or exotic scenarios.
We also include in these graphs contours of the different values of $\Omega
_{s}$, given by Eq. (\ref{omega}). One can readily see that the flux of 511
KeV photons is a very powerful tool for probing sterile neutrinos, as even
minuscule abundances of sterile neutrinos, $\Omega_s\sim 10^{-9}$, can give
observable fluxes.

To produce a model-independent exclusion plot, we assume the conservative
value of $\alpha =0.2$ since this value of $\alpha $ yields the lowest count,
and we allow $|V|$ to vary. Then the conservatively excluded region will be the
overlap of the two $\alpha =0.2$ plots for the extreme values of $|V|=0,1$,
shown in Fig. 2. We can also disconnect the result from a particular
functional form of the dark-matter density near the center by fixing its
value at $r=3\,\,$kpc and keeping it constant inside that radius while
keeping the same parametrization (\ref{newprof}) with $\alpha =0.2$ outside
of this region. Clearly, this will yield the most conservative outcome, and
we show it also in Fig. 2. Notice that the assumption about the Majorana nature 
of sterile neutrinos would increase all rates by a factor of 2, and thus lead to 
a larger excluded region.

\begin{figure}[tbp]
\centerline{\psfig{file=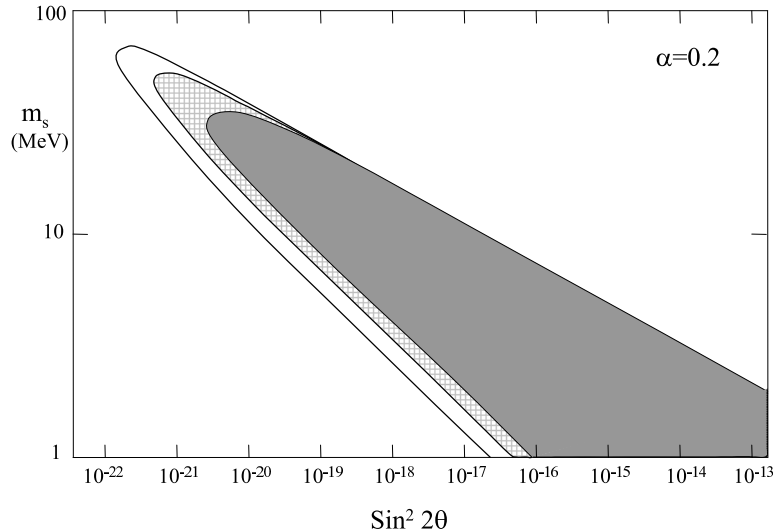,width=8cm,angle=-90}}

\vspace{1cm}

\caption{The exclusion plot for sterile neutrino parameters based on
conservative assumptions about the dark matter profile near the center of the
Galaxy. The unshaded area corresponds to $|V|=1$ and the hatched area to $|V|$=0. 
The solid area is the most conservative model-independent result, corresponding to 
a dark matter density which is constant inside 3 kpc. }
\end{figure}

Returning to Fig. 1, we observe that the combinations of $m_s$ and $\theta$
that lie at the boundary of the shaded region give the photon flux that
can explain the INTEGRAL/SPI signal. We see that the observed flux can be
accommodated with densities in a wide range of sterile neutrino abundances, $%
10^{-9}<\Omega <10^{-5}$. Each of these curves contains two branches: for a
specified value of the mixing angle $\theta$ there are two values of $m_s$
consistent with the photon flux. The branch with lower $m_s$ corresponds to
the sterile neutrino lifetimes that are larger than the age of the Universe, 
$\Gamma_N\tau <1$, and low values of $\Omega_s$ in this case are caused by
rather inefficient production of sterile neutrinos in the hot Universe. Such
low $\Omega_s$ and mixing angles makes this scenario immune to other
possible constraints. At the same time, for the upper branch of the curve, $%
\Gamma_N \tau >1$ and small $\Omega_s$ can also be caused by the exponential
suppression in (\ref{omega}). Consequently, for $m_s \sim$ few MeV along
this branch the energy density of sterile neutrinos at higher redshifts
could be of order one compared to the rest of matter, and thus is subject to
other cosmological constraints such as diffuse photon background,
(see {\em e.g.} \cite{DZ,AFP}). 

If we assume that the flux of photons seen by INTEGRAL/SPI comes from the
decay of sterile neutrinos, we can use the angular distribution to determine
the preferred cold dark matter density profile in the inner galactic region
in a similar fashion as in Ref. \cite{Silk}. Averaging over the angular
resolution of the spectrometer and using the common flux normalization at
the zero angle result in the angular profiles shown in Fig.3 for $\alpha
=0.1 $ and $\alpha =0.2$. The distribution for $\alpha =0.2$ falls well
within the 2$\sigma $ confidence interval, and thus is clearly favored by
the data over smaller values of $\alpha $. The angular distribution that we
calculated with the power-like density function $\rho (r)\varpropto
1/r^{\gamma }$ for $\gamma =1$ comes very close to the one for $\alpha =0.2.$
\begin{figure}[tbp]
\centerline{\psfig{file=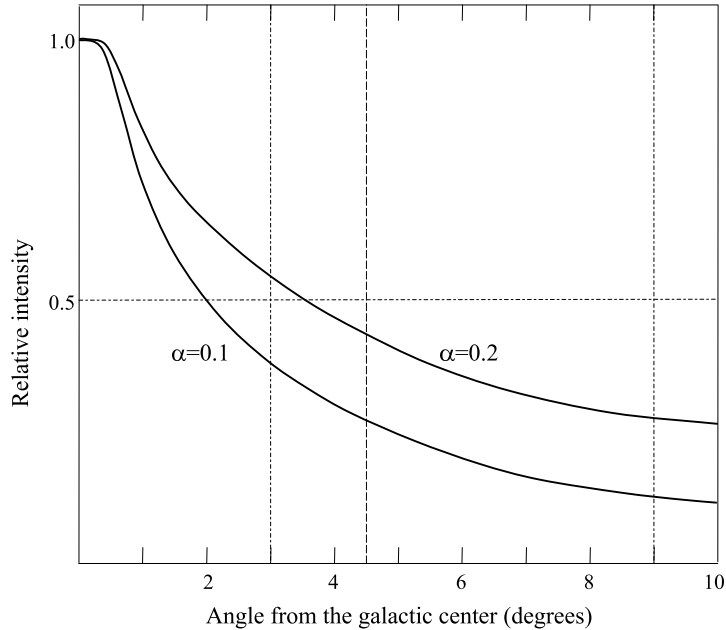,width=8cm,angle=-90}}

\vspace{1cm}
\caption{The angular distribution of 511 KeV photons for two different inner
halo profiles, $\alpha =0.1$ and $0.2$, averaged over the 2$^{{\rm {o}}}$ 
resolution of the 
SPI spectrometer. The experimental data were fit with a gaussian 
of full width at half maximum at a 9$^{{\rm {o}}}$ diameter circle 
(shown as a vertical dashed line),
with a $2\sigma$ confidence interval of $6^{{\rm {o}}}-18^{{\rm {o}}}$
(shown as vertical dashed-dotted lines). As can be seen, 
the distribution for $\alpha =0.2$ falls well within the 2$\sigma$ confidence 
interval.  }
\end{figure}

The photon width $\Gamma _{N\to \nu \gamma }$, although smaller than $\Gamma
_{N\to \nu e\bar{e}}$, should lead to the additional photon line at $%
E_\gamma = m_s/2$. The intensity of $m_s/2$ line, can nevertheless come
close to that of 511 KeV if $m_s$ is large because of $m_s/(2m_e)$
enhancement factor. Unfortunately, the prediction of photon fluxes at $m_s/2$
relative to 511 KeV, 
\begin{equation}
\frac{\Phi_{0.5 m_s\gamma}}{\Phi_{511\gamma}}= \frac{\Gamma _{N\to \nu
\gamma }}{2\Gamma _{N\to \nu e\bar{e}}} = \frac{27\alpha}{\pi(4|V|^2+1)} = 
\frac{0.031}{4|V|^2+1},
\end{equation}
varies by a factor of 5 depending on $|V|$.

It is important to compare the limits from positron production in 
the decays of the sterile neutrinos  with the limits derived from a 
comparison to the cosmological diffuse gamma background. 
Ref. \cite{DH} estimates the diffuse photon background produced by the
radiative sterile neutrinos with an additional assumpiton of 
sterile neutrinos constituting the dominant part of dark matter. 
As a a first approach, we can effectively compare their results by replacing their
assumed neutrino abundance of 0.3 by our calculated $10^{-9}\lesssim \Omega
_{\text{sterile}}\lesssim 10^{-5}$, which relaxes their constraints on $%
\sin^{2}\theta $ by many orders of magnitude. For the diffuse photon
background this meant, effectively, a bound on the lifetime $\tau >4\times
10^{15}$ years, but as can be seen from our Eq. \ref{generic} our
sensitivity to $\tau $ is at the level of $5\times 10^{17}$ years asuming
the same abundance, 1 MeV neutrinos and a minimum branching ratio to $%
e^{+}e^{-}$ of 0.1. Therefore, it appears that 511 KeV photon flux provides more
stringent bounds than the diffuse photon background by one to two orders of
magnitude. It is nevertheless important to consider this issue in more detail, 
as the experimental data on photon background used in Ref. \cite{DH} 
are clearly outdated. 

The spectral density of the photon flux can be expressed in terms 
of the sterile neutrino mass $m_s$, the branching ratio to photons, the neutrino 
lifetime, the "initial" neutrino abundance $\Omega_{0s}$, 
the age of the Universe $\tau$ and its energy density today $\rho_0$,
\begin{equation}
\fr{d\Phi_\gamma}{dE}=\fr{1}{4\pi}\Omega_{0s}\rho_0 {\rm Br}_{\gamma}\fr 3m 
\left (\fr{2E}{m}\right)^{1/2}\Gamma_{N}\tau\exp\{-\Gamma_N\tau(2E/m)^{3/2}\}.
\label{photonspectrum}
\end{equation}
The maximum in this spectrum corresponds to photon energies
\be
E_{\rm max} = \left\{\begin{array}{c} m/2~~~ {\rm for}~~~   3\Gamma_n \tau<1,\\
(3\Gamma_n \tau)^{-1}m/2~~~ {\rm for}~~~   3\Gamma_n \tau>1\end{array}\right.
\label{Emax}
\ee
On the experimental side, \cite{EGRET-COMPTEL}, we assume a conservative bound on 
the spectrum of the cosmological diffuse gamma ray background,
\be
\left(\fr{d\Phi_\gamma}{dE}\right)_{\rm exp} \la \fr{10^{-2}}{E^2} 
~{\rm MeV~str^{-1}cm^{-2}s^{-1}}
\ee
Plugging in the expression for $E_{\rm max}$ (\ref{Emax}) to the photon 
spectrum (\ref{photonspectrum}), we require 
$(d\Phi_\gamma/dE)_{\rm max}$ to be smaller than 
 $(d\Phi_\gamma/dE)_{\rm exp}$. This results in the following 
constraint on the parameters of the model,
\be
3.5\times 10^{-7}> \left\{\begin{array}{c}\fr{13}{9+4|V|^2}
\Omega_{0s}h^2 x\exp\{-\fr 13 x\}~~~ {\rm for}~~~   x<1,\\\,\\
\fr{13}{9+4|V|^2}\Omega_{0s}h^2 x^{-3/2}\exp\{-\fr {1}{3} x^{-1/2}\}~~~ {\rm for}~~~   x>1,
\end{array}\right.
\ee
where $x = 3 \Gamma_N \tau$ is introduced for concision.  
This constraint simplifies when $x\ll 1$, in which case we obtain 
a very powerful limit on the combination of masses and mixing angles,
\be
\left(\fr{\sin^22\theta}{10^{-19}}\right)^2
\left(\fr{m_s}{\rm 1~ MeV}\right)^5 \la 0.8
\ee
where $V$ is taken to be one. 

Plotting this constraint in Figure 1, we superimpose it on the 
limits obtained from the 511 KeV line. We observe that the 511 KeV line typically 
provides better sensitivity to parameters of the model when $\Gamma_N \tau <1$, 
while the cosmological background of gamma can provide more restrictive bounds 
if $\Gamma_N \tau >1$.
Unlike the limit from 511 KeV photons, 
the cosmological background constraint is not sensitive 
to the details of the dark matter distribution in the galactic center.

\section{Positron flux from decaying dark matter}

In this section, we generalize the analysis of the previous section to an
arbitrary decaying relic, without immediately specifying its cosmological
origin and keeping its abundance $\Omega _{S}$ as a free parameter, $\Omega
_{S}\leq \Omega _{DM}$. The equality would mean that this relic is the main
component of the cold dark matter. The flux of the galactic 511 KeV photons,
normalized to the measured value, can be written as a function of the mass $%
m_{S}$, the decay width into $e^{+}e^{-}$ and $\Omega _{S}$: 
\begin{equation}
\frac{\Phi (m_{s},\Gamma _{e^{+}e^{-}},\Omega _{s})}{\Phi _{{\rm exp}}}%
\simeq (5\times 10^{17}{\rm yr}\times \Gamma _{e^{+}e^{-}})~\frac{10{\rm MeV}}{m_{S}}%
~\frac{\Omega _{S}}{0.23}.  \label{generic}
\end{equation}
In this equation, $\alpha =0.2$ was used. We observe that even if dark
matter particles have much longer lifetimes than the age of the Universe,
they are still capable of providing significant numbers of positrons.

Sterile neutrinos in the mass range of 1 MeV -- 50 MeV can account for the
galactic positrons but cannot constitute any appreciable fraction of the
dark matter if they are produced thermally from active neutrino species
under the assumption of a small lepton asymmetry. To this end, it is
interesting to relax the assumptions that led to Eq. (\ref{omega}) and using
Eqs. (\ref{generic}) and (\ref{e+e-}) estimate the resulting photon flux, 
\begin{equation}
\frac{\Phi (m_{s},\theta ,\Omega _{s})}{\Phi _{{\rm exp}}}\simeq
1.7~(1+4|V|^{2})~\frac{\sin ^{2}2\theta }{10^{-24}}~\left( \frac{m_{s}}{10~%
{\rm MeV}}\right) ^{4}~\frac{\Omega _{s}}{0.23}.  \label{relax}
\end{equation}
Combining the analysis of Ref. \cite{AFP} at $|V|=1$ and the lepton
asymmetry $L=0.1$ with Eq. (\ref{relax}), we find that all points on the $%
\theta -m_{s}$ parameter space that correspond to $\Omega _{s}\ge 10^{-3}$
are firmly ruled out by the excessive photon flux. Therefore, even for an
arbitrary lepton asymmetry, sterile neutrinos with $m_{s}>1$ MeV cannot
constitute a significant fraction of dark matter. The combination of both features,
the sterile neutrino dark matter ($m_{s}>1$MeV) and acceptable positron
numbers from its decay, may only come at the price of abandoning the
conventional mechanism of thermal transfer of active neutrinos into sterile,
and assuming some initial sterile neutrino energy density due to either
non-thermal production after inflation and/or decay of some heavy non-SM
particles into $N$ before the electroweak epoch.

Another example of an unstable relic that could produce positrons is an
extremely weakly interacting scalar or pseudoscalar particle $\phi $ in the
same MeV mass range. Quite generically, we can write its interaction with
operators composed from the Standard Model fields as 
\begin{equation}
{\cal L}_{\phi }=\frac{1}{2}\partial _{\mu }\phi \partial ^{\mu }\phi -\frac{%
1}{2}m_{\phi }^{2}\phi ^{2}+\frac{\phi }{M_{*}}\sum_{i}c_{i}O_{i}^{(4)},
\label{scalar}
\end{equation}
where the leading order dimension 4 operators that induce the decay of $\phi 
$ are included. In models with extra space-like dimensions, a scalar
particle called radion would couple to the trace of the stress-energy tensor
of visible matter. If $\phi $ represents a radion, a scalar mode
corresponding to the fluctuation of the volume of extra dimension, the
coupling constant $M_{*}$ can vary within a large range, from $M_{{\rm pl}}$ or
even larger scale down to $M_{W}$ with the latter being the case in the
Randall-Sundrum model \cite{RS}.

For 1 MeV$<m_{\phi }\mathrel{\raise.3ex\hbox{$<$\kern-.75em\lower1ex\hbox{$%
\sim$}}}$ 100 MeV, the leading interaction with fermions is given by $\phi
M_{*}^{-1}m_{e}\bar{e}e$ for scalars and $\phi M_{*}^{-1}m_{e}\bar{e}i\gamma
_{5}e$ for pseudoscalars. In both cases, the width to $e^{+}e^{-}$ is given
by 
\begin{equation}
\Gamma _{\phi \to e^{+}e^{-}}=\frac{m_{e}^{2}m_{\phi }}{8\pi M_{*}^{2}}%
\simeq (2\times 10^{17}~{\rm yr})^{-1}\frac{m_{\phi }}{1~{\rm MeV}}\,\left( 
\frac{10^{19}~{\rm GeV}}{M_{*}}\right) ^{2},  \label{gammaphi}
\end{equation}
where the threshold effects are neglected, $2m_{e}<m_{\phi }$. Whether Eq. (%
\ref{gammaphi}) provides the dominant contribution to the total decay width
depends on whether or not the anomalous interaction with photons, $\phi
F_{\mu \nu }F^{\mu \nu }$ or $\phi F_{\mu \nu }\tilde{F}^{\mu \nu }$, is
generated at one loop. This depends on the UV completion of the {\em 
effective} Lagrangian (\ref{scalar}), and both outcomes, zero or non-zero couplings to 
photons, are possible (for the scalar
case see {\em e.g.} \cite{OP} and references therein). In particular, the
one-loop anomalous coupling of scalars and photons is not generated in the
models where the interaction term of the scalar $\phi $ with matter is
obtained from a conformal rescaling of Brans-Dicke theory with the mass term
for $\phi $. If $\phi F_{\mu \nu }^{2}$ arises at two loops or not at all,
then the total decay width is dominated by $\Gamma _{\phi \to e^{+}e^{-}}$.
For the pseudoscalar case, one could also avoid couplings to photons by
postulating a derivative interaction with the SM fields, {\em i.e.} $\partial _{\mu }\phi
\bar \psi\gamma _{\mu }\gamma _{5}\psi$. Taking the point of view that $e^{+}e^{-}$
represents the main decay channel, we combine Eqs. (\ref{generic}) and (\ref
{gammaphi}) to provide an estimate of the photon flux from the Galactic
bulge, 
\begin{equation}
\frac{\Phi (M_{*},\Omega _{\phi })}{\Phi _{{\rm exp}}}\simeq 25~\left( \frac{%
10^{19}~{\rm GeV}}{M_{*}}\right) ^{2}~\frac{\Omega _{\phi }}{0.23},
\end{equation}
which turns out to be independent of $m_{\phi }$. It is remarkable that $%
M_{*}$ as large as the Planck scale, 
\begin{equation}
M_{*}\sim 5\times 10^{19}~{\rm GeV}\sqrt{\Omega _{\phi }/0.23},
\label{mstar}
\end{equation}
can be probed this way for maximal $\Omega _{\phi }$!

Is it realistic to have $\Omega _{\phi }\sim 0.23$? There are many possible
scenarios of non-thermal production of scalars and pseudoscalars. In fact,
the scalar field energy density can be over-produced easily leading to the
closure of the Universe (moduli problem). This is, of course, very much
model-dependent, and can be easily avoided in some scenarios. Thermal
production can also be quite efficient if in addition to very weak decay terms the
scalar $\phi $ has appreciable bilinear couplings to the Standard Model
fields. For example, the Lagrangian \cite{MD1,BPtV,MD2}, 
\begin{equation}
{\cal L}_{\phi }=\frac{1}{2}\partial _{\mu }\phi \partial ^{\mu }\phi -\frac{%
1}{2}(m_{0}^{2}\phi ^{2}-\lambda \phi ^{2}H^{\dagger }H)+\frac{\phi }{M_{*}}%
m_{e}\bar{e}e,  \label{phiagain}
\end{equation}
where $H$ is the Standard Model Higgs doublet, has the potential of
providing a dark matter candidate $\phi $. To ensure the right abundance of $%
\phi $ one can choose $\lambda $ to be large, in which case $\phi $ is
thermalized above $m_{\phi }$ and depletes its abundance via annihilation at
the freeze-out. In this case however, there is a direct experimental bound
on $m_{\phi }$ that requires it to be heavier than 400 MeV \cite{victoria},
thus making it unsuitable for explaining the positron flux. Another
possibility for obtaining a cosmologically required value of $\Omega _{\phi
} $ is to assume a small value of $\lambda $, initial absence of $\phi $
scalars, and their thermal production at $T\sim M_{W}$ \cite{MD2}. This
option is indeed viable, and $\lambda \sim 10^{-7}$ would lead to $\Omega
_{\phi }\sim 0.2$. Thus, we see that the scalar field model (\ref{phiagain})
can account for the existence of cold dark matter and lead to the observable
fluxes if condition (\ref{mstar}) is satisfied. In the original
formulation of this model without the decay term the stability of the scalar
was ascribed to either $Z_{2}$ symmetry \cite{BPtV} or global $U(1)$
symmetry if $\phi $ is a complex scalar \cite{MD1}. The appearance of the
decay terms $\sim M_{{\rm Pl}}^{-1}\phi m_{e}\bar{e}e$ might be speculated
to arise due to the violation of these symmetries in string theory/quantum
gravity at the Planck scale. 

\section{Discussion and Conclusions}

We have shown that the decaying relics that may or may not constitute the
main component of dark matter can easily account for the flux of galactic
photons observed by the INTEGRAL/SPI instrument \cite{SPI}. The lifetimes of
these relics of order $10^{17}$ yr, {\em i.e.} much larger than the age of
the Universe, can be probed this way. This is also longer than the 
lifetime of unstable $\sim$MeV-scale relics that can be probed with 
diffuse cosmic gamma-ray background \cite{DH}. This sensitivity 
motivates further theoretical studies of the flux of 511 KeV photons from the 
bulge.

We have studied in detail the model of sterile neutrinos produced in the
early Universe by conversion from active neutrino species. We have shown
that if the decay to $e^{+}e^{-}$ pairs is kinematically allowed, the
branching ratio in this mode is never smaller than 10\% even if the sterile
neutrinos have no charged currents with electrons ($V=0$). Using previous
results of abundance calculation \cite{DW,AFP} and  the total decay width of the
sterile neutrino and its branching to positrons obtained in this work, 
we were able to calculate
the flux of the 511 KeV photons as a function of the mixing angle $\theta $
with active species and the sterile neutrino mass $m_{s}$. Over a large
domain of the parameter space for 1 MeV$<m_{s}\mathrel{\raise.3ex%
\hbox{$<$\kern-.75em\lower1ex\hbox{$\sim$}}}$50 MeV this flux turns out to
be significantly larger than the observed value. This enables us to exclude
a significant portion of the parameter space of sterile neutrinos regardless
of whether the source of the galactic positrons is astrophysical or it is due to exotic
physics. In particular, we are able to rule out any cosmologically
significant quantities of sterile neutrinos, $\Omega _{s}<10^{-3}$ with
masses larger than 1 MeV.

The boundary of the excluded region corresponds to fluxes that are
comparable to the observed value. If indeed the experimental results can be
interpreted as the consequence of positrons from the decaying sterile
neutrinos, the photon flux points to a sterile neutrino abundance in the
range of $10^{-9}\mathrel{\raise.3ex\hbox{$<$\kern-.75em\lower1ex\hbox{$%
\sim$}}}\Omega _{{\rm sterile}}\mathrel{\raise.3ex\hbox{$<$\kern-.75em%
\lower1ex\hbox{$\sim$}}}10^{-5}$. It is also interesting to note that the
upper value for the mass of sterile neutrinos ($\sim $50 MeV) compatible
with the observed flux of photons comes naturally into this analysis, and
not as an external requirement as in the case of annihilating dark matter 
\cite{Silk}. The radial distribution of the decaying relics inside the
galactic halo should trace the distribution of the dark matter. Thus, if
indeed future data support the link of galactic positrons to exotic
physics ({\em e.g.} by firmly establishing their diffuse origin as opposed
to point-like sources), the two scenarios, decaying relics and annihilating
dark matter, can be distinguished by the angular distribution of the signal.
The signal from decaying relics is proportional to the number density of the
dark matter particles $n(r)$ while the annihilating dark matter has $%
n^{2}(r) $ dependence. Thus, the decaying relics would suggest steeper
Galaxy profiles than the scenario with the annihilation of dark matter. In
both scenarios, decaying relics and annihilating dark matter, one would
generally expect the additional sharp line due to the annihilation or decay
to photons. In the case of decaying sterile neutrinos, the flux of photons
with energies of $0.5m_{s}$ is predicted to be at the level of few percent
of the flux of 511 KeV neutrinos. Finally, we would like to comment that the
scenario with annihilating dark matter can be searched for experimentally
using the existing data on $e^{+}e^{-}$ colliders. Since the results of Ref. 
\cite{Silk} suggest an appreciable low-energy dark matter annihilation cross
section into $e^{+}e^{-}$, one would naturally look for the missing energy
signal (plus a hard photon) in the $e^{+}e^{-}$ collisions at higher
energies where the SM model cross sections become smaller and while the
cross section of the MeV-mass dark matter production would become larger.

\section*{Acknowledgements}

We would like to thank I. Mocioiu,
J. Navarro and A. Ritz for useful conversations, as well as S. Hansen and S. Sarkar 
for pointing out additional important references on the subject. 
The work of M.P. is partially supported 
by NSERC of Canada and PPARC of the UK.



\begin{thebibliography}{99}
\bibitem{SPI}  P.~Jean {\it et al.}, 
Astron.\ Astrophys.\ {\bf 407} (2003) L55; J.~Knodlseder {\it et al.}, 
arXiv:astro-ph/0309442.

\bibitem{positrons}  M. Leventhal {\em et al.}, Astrophys.\ J.\ Lett. 25
(1993) 405; D. M. Smith {\em et al.}, Astrophys.\ J.\ 414 (1993) 165; M. J.
Harris {\em et al.}, Astrophys.\ J.\ Lett. 55 (1998) 501.

\bibitem{Silk}  C.~Boehm, D.~Hooper, J.~Silk and M.~Casse, 
arXiv:astro-ph/0309686.

\bibitem{NFW}  J.~F.~Navarro, C.~S.~Frenk and S.~D.~M.~White, 
Astrophys.\ J.\ {\bf 462} (1996) 563.

\bibitem{BPtV}  C.~P.~Burgess, M.~Pospelov and T.~ter Veldhuis, 
Nucl.\ Phys.\ B {\bf 619} (2001) 709.

\bibitem{Fayet}  C.~Boehm, P.~Fayet and J.~Silk, 
arXiv:hep-ph/0311143.

\bibitem{DW}  S.~Dodelson and L.~M.~Widrow, 
Phys.\ Rev.\ Lett.\ {\bf 72} (1994) 17.

\bibitem{DHRS} A.D. Dolgov, S.H. Hansen, G. Raffelt, D.V. Semikoz,
Nucl. Phys. B590 (2000) 562.

\bibitem{DH} A.D. Dolgov, S.H. Hansen,
Astroparticle Physics 16 (2001) 339. 

\bibitem{AFP}  K.~Abazajian, G.~M.~Fuller and M.~Patel, 
Phys.\ Rev.\ D {\bf 64} (2001) 023501.

\bibitem{Irina}G.~M.~Fuller, A.~Kusenko, I.~Mocioiu and S.~Pascoli,
Phys.\ Rev.\ D {\bf 68} (2003) 103002.


\bibitem{BPS}  V.~D.~Barger, R.~J.~N.~Phillips and S.~Sarkar, Phys.\ Lett.\
B {\bf 352} (1995) 365 [Erratum-ibid.\ B {\bf 356} (1995) 617].

\bibitem{WP}  P.~B.~Pal and L.~Wolfenstein, 
Phys.\ Rev.\ D {\bf 25} (1982) 766.

\bibitem{Navarro}  J.~F.~Navarro {\it et al.}, arXiv:astro-ph/0311231.

\bibitem{DZ}  A.~D.~Dolgov and Y.~B.~Zeldovich, 
Rev.\ Mod.\ Phys.\ {\bf 53} (1981) 1.

\bibitem{EGRET-COMPTEL} P.~Sreekumar {\it et al.}  [EGRET Collaboration],
Astrophys.\ J.\  {\bf 494} (1998) 523.

\bibitem{RS}  L.~Randall and R.~Sundrum, 
Phys.\ Rev.\ Lett.\ {\bf 83} (1999) 3370.

\bibitem{OP}  K.~A.~Olive and M.~Pospelov, 
Phys.\ Rev.\ D {\bf 65} (2002) 085044.

\bibitem{MD1}  J.~McDonald, 
Phys.\ Rev.\ D {\bf 50} (1994) 3637.

\bibitem{MD2}  J.~McDonald, 
Phys.\ Rev.\ Lett.\ {\bf 88} (2002) 091304.

\bibitem{victoria}  C.~Bird, P.~Jackson, R.~Kowalewski and M.~Pospelov, 
arXiv:hep-ph/0401195, to appear in PRL.
\end{thebibliography}
\end{document}